\documentclass[
    trackchanges,
    twocolumn,
    twocolappendix,
]{aastex701}
\usepackage{amsmath}
\usepackage{graphicx}
\usepackage{dcolumn}
\usepackage{bm}
\usepackage{siunitx}
\usepackage{braket}
\usepackage{comment}
\usepackage{wasysym} 

\newcommand\orcid[1]{\href{https://orcid.org/#1}{\includegraphics[height=9pt]{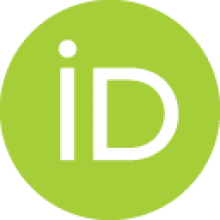}}}
\newcommand{\change}[1]{#1}

\begin{document}


\title{The Simons Observatory: Rotation Performance of Cryogenic Half-Wave Plate Polarization Modulators}

\author{Kyohei~Yamada~\orcid{0000-0003-0221-2130}}
\affiliation{Joseph Henry Laboratories of Physics, Jadwin Hall, Princeton University, Princeton, NJ 08544, USA\looseness=-1}
\email{kyoheiy@princeton.edu} 

\author{Bryce~Bixler~\orcid{0009-0008-4312-6814}}
\affiliation{Department of Physics, University of California, San Diego, La Jolla, CA 92093, USA}
\email{bbixler@ucsd.edu}

\author{Junna~Sugiyama~\orcid{0009-0007-7435-9082}}
\affiliation{Department of Physics, Graduate School of Science, The University of Tokyo, Tokyo 113-0033, Japan\looseness=-1}
\email{sugiyama-junna605@g.ecc.u-tokyo.ac.jp}

\author{Daichi~Sasaki~\orcid{0009-0003-2513-2608}}
\affiliation{Department of Physics, Graduate School of Science, The University of Tokyo, Tokyo 113-0033, Japan\looseness=-1}
\email{daichi.sasaki@phys.s.u-tokyo.ac.jp}

\author{Yuki~Sakurai~\orcid{0000-0001-6389-0117}}
\affiliation{Department of Mechanical and Electrical Engineering, The Suwa University of Science, Nagano 391-0213, Japan\looseness=-1}
\affiliation{Kavli Institute for the Physics and Mathematics of the Universe (WPI), UTIAS, The University of Tokyo, Chiba 277-8583, Japan\looseness=-1}
\email{sakurai_yuki@rs.sus.ac.jp}

\author{Kam~Arnold~\orcid{0000-0002-3407-5305}}
\affiliation{Department of Astronomy and Astrophysics, University of California San Diego, USA}
\affiliation{Department of Physics, University of California, San Diego, La Jolla, CA 92093, USA}
\email{arnold@ucsd.edu}

\author{Samuel~Day-Weiss~\orcid{0009-0003-5814-2087}}
\affiliation{Joseph Henry Laboratories of Physics, Jadwin Hall, Princeton University, Princeton, NJ 08544, USA\looseness=-1}
\email{dayweiss@princeton.edu}

\author{Nicholas~Galitzki~\orcid{0000-0001-7225-6679}}
\affiliation{Department of Physics, University of Texas at Austin, Austin, TX 78712, USA\looseness=-1}
\affiliation{Weinberg Institute for Theoretical Physics, Texas Center for Cosmology and Astroparticle Physics, Austin, TX 78712, USA\looseness=-1}
\email{nicholas.galitzki@austin.utexas.edu}

\author{Bradley~R.~Johnson~\orcid{0000-0002-9828-3525}}
\affiliation{University of Virginia, Department of Astronomy, Charlottesville, VA 22904, USA\looseness=-1}
\email{bradley.johnson@virginia.edu}

\author{Akito~Kusaka~\orcid{0009-0004-9631-2451}}
\affiliation{Department of Physics, Graduate School of Science, The University of Tokyo, Tokyo 113-0033, Japan\looseness=-1}
\affiliation{Kavli Institute for the Physics and Mathematics of the Universe (WPI), UTIAS, The University of Tokyo, Chiba 277-8583, Japan\looseness=-1}
\affiliation{Research Center for the Early Universe, School of Science, The University of Tokyo, Tokyo 113-0033, Japan}
\affiliation{Physics Division, Lawrence Berkeley National Laboratory, Berkeley, CA 94720, USA}
\email{akusaka@phys.s.u-tokyo.ac.jp}

\author{Lyman~A.~Page~\orcid{0000-0002-9828-3525}}
\affiliation{Joseph Henry Laboratories of Physics, Jadwin Hall, Princeton University, Princeton, NJ 08544, USA\looseness=-1}
\email{page@princeton.edu}

\author{Yoshinori~Sueno~\orcid{0000-0002-3644-2009}}
\affiliation{Joseph Henry Laboratories of Physics, Jadwin Hall, Princeton University, Princeton, NJ 08544, USA\looseness=-1}
\email{yoshinorisueno@gmail.com}

\author{Satoru~Takakura~\orcid{0000-0001-9461-7519}}
\affiliation{Department of Physics, Graduate School of Science, The University of Tokyo, Tokyo 113-0033, Japan\looseness=-1}
\email{satoru.takakura@phys.s.u-tokyo.ac.jp}



\begin{abstract}
We present the on-site rotation performance of the first three cryogenic continuously rotating half-wave plate (HWP) polarization modulators for the Simons Observatory small aperture telescopes (SATs). The SATs operate at an altitude of 5200\,m in the Atacama Desert in northern Chile, and measure the degree-scale cosmic microwave background polarization to search for primordial $B$-mode polarization. To this end, the SATs employ a 505\,mm diameter 50\,K cryogenic HWP polarization modulator to suppress atmospheric $1/f$ noise and to mitigate systematic uncertainties. We present methods for reconstructing the rotation of our polarization modulators, enabling detailed evaluations of their rotation angle accuracy, stability, displacements, and vibrations. We achieve the required rotation angle accuracy in more than 99.9\% of observations, with a median noise level of 0.16\,$\mu$rad$\sqrt{\text{s}}$. We also achieve one-dimensional measurement of the rotor displacement with an accuracy of $0.06\,\mu\text{m}\sqrt{\text{s}}$. Our results demonstrate the on-site rotation performance, rotation angle reconstruction method, and robustness of the polarization modulators. This is a crucial step towards achieving the SAT science goals.
\end{abstract}

\section{Introduction} \label{sec:intro}

A frontier of cosmic microwave background (CMB) research is the measurement of its polarization. In particular, the large angular scale parity-odd $B$-mode polarization is sensitive to primordial gravitational waves that could be produced by cosmic inflation \citep{kamionkowski_probe_1997, seljak_signature_1997}. The amplitude of primordial $B$-mode polarization is parametrized by the ratio of the primordial tensor and scalar perturbations, $r$ \citep{1967ApJ}.
The Simons Observatory is an ongoing CMB experiment that started observations in 2023 and is located at an altitude of 5200\,m in the Atacama Desert of northern Chile \citep{so_science_goals}. To date, the observatory includes three small aperture telescopes (SATs), each with a 42\,cm aperture and 12,348 transition edge sensor detectors at maximum.
With a five-year survey, the three SATs aim to measure degree-scale CMB polarization to detect or constrain primordial $B$-mode polarization with the goal sensitivity of $\sigma(r) = 0.002$ \citep{so_science_goals}. In the near future there will be an expansion of the observatory with three additional SATs \citep{so_2025_sat_expansion}.

Rapid modulation of polarization is one of the key techniques for precise measurement at large angular scales, separating polarization signal from unpolarized atmospheric $1/f$ noise and suppressing instrumental systematic effects \citep{POLAR_2003, DASI_2005, Barkats_2005, matsumura_phd, johnson_maxipol_2007, Chen_2009, klein_cryogenic_2011, QUIET_2012, Moyerman_2013, kusaka_modulation_2014, Miller_2016, Hill2016,  takakura_performance_2017, johnson_large-diameter_2017, HillPB2bCHWP2020, LB_PMU_LF, GB2020, Harrington_2021, class_hwp_2024, BixlerPb2bCHWP2026}.
The SATs employ a 505\,mm-diameter 50\,K cryogenic continuously rotating half-wave plate (HWP) as a polarization modulator \citep{SO_CHWP, SO_AHWP, Sasaki_2025}. \change{The HWP is a birefringent plate, and the polarization angle of incident linearly polarized radiation is mirrored about the fast axis upon exit.} We achieve the polarization modulation at 8\,Hz by continuously rotating the HWP at 2\,Hz. Its rotation mechanism consists of a contactless superconducting magnetic bearing (SMB) and synchronous electromagnetic motor; its rotation is monitored by a pair of optical encoders \citep{SO_CHWP}.  

For precise polarimetry using modulation techniques, a detailed characterization of the modulation, i.e., mechanical motion of the HWP, is crucial. This is particularly important for the modern millimeter-wave instruments that employ large arrays of detectors, because a misestimation of the modulation could lead to largely coherent systematic bias in the detector data. However, characterizing the motions of magnetically levitating and spinning object in a cryogenic vacuum space poses challenges, and various techniques have been developed \citep{Hull_2005, Sakurai_2017, Sakurai_rs_2018, Sugiyama:2022flm}. In this paper, we describe advancing the encoder data analysis techniques to address these challenges.

To date, the Simons Observatory has deployed two SATs with mid-frequency bands (MF: 90/150\,GHz) and one with ultra-high-frequency bands (UHF: 220/280\,GHz). We denote the two MF-SATs as SAT1 and SAT3 and the UHF-SAT as SAT2. While the design of HWP is different for MF and UHF, the design of its rotation mechanism is identical.
We evaluate the accuracy of the HWP angle reconstruction, and present the rotation performance of the HWP polarization modulators in three SATs from the initial year of observations. 
The organization of this paper is as follows: Section~\ref{sec:method} provides an overview of the HWP polarization modulator, especially its rotation mechanism and the rotation angle reconstruction method.
Section~\ref{sec:results} describes the in field performance.
Section~\ref{sec:conclusion} describes the conclusion.

\section{Methods} \label{sec:method} 
The SATs of the Simons Observatory, designed with refractive optics, employ a cryogenic continuously rotating HWP as a polarization modulator \citep{Galitzki_2024}. The incident radiation modulated by the rotating HWP is then focused by the 1\,K optics tube with silicon lenses. It is detected by the array of transition edge sensors on the focal plane at $\leq$100\,mK \citep{Stevens2020}. The detector signals are read out by SMuRF electronics and are recorded at the rate of 200\,Hz \citep{10.1117/12.2314435}. 


\subsection{Overview of SAT HWP polarization modulator}
\begin{figure}
    \centering
    \includegraphics[width = 0.48\textwidth]{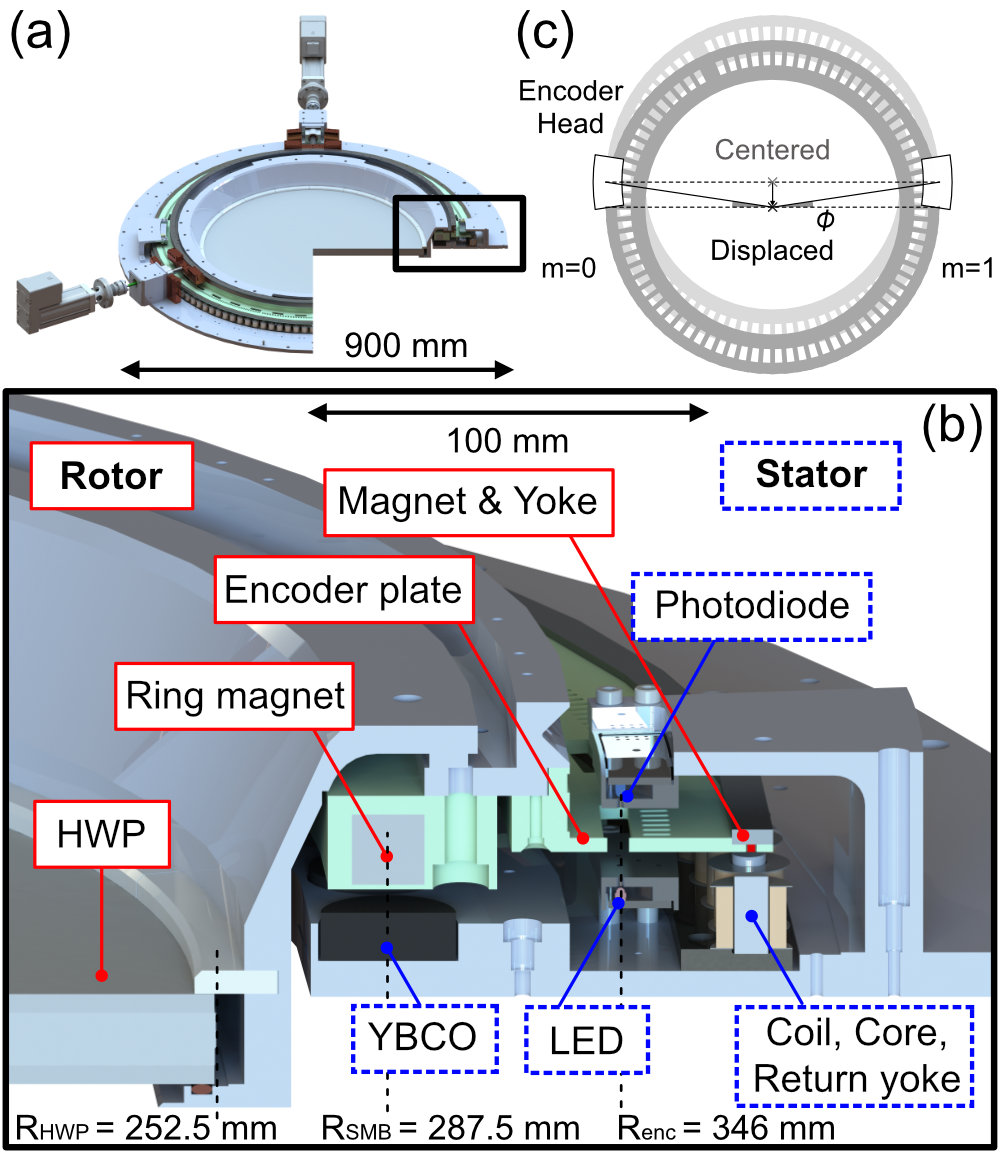}
    \caption{HWP polarization modulator. (a) CAD view of the whole system. (b) Magnified cross sectional view of the rotation mechanism. The rotating components are labeled with red boxes, and the stationary components are labeled with blue dotted boxes. (c) Schematic view of the rotor and encoder heads when the boresight angle of the telescope is zero. When the rotor is off-centered perpendicular to the line connecting the encoders, the encoder incorrectly measures the rotation angle by $\phi$.}
    \label{fig:method}
\end{figure}

Figure~\ref{fig:method} shows the rotation mechanism of the HWP polarization modulator of the SAT \citep{SO_CHWP}. The mechanism is composed of rotor and stator sections held together through a SMB consisting of ring-shaped yttrium barium copper oxide (YBCO) superconductor assembly and ring-shaped neodymium magnet assembly. Rotation is achieved by alternating currents through the stator solenoid coils, magnetically coupling them to embedded sprocket neodymium magnets on the outer rim of the rotor's G10 encoder plate. We employ two optical encoder assemblies that are installed 180$^\circ$ from each other, each with a set of five photo-diodes mounted on an arm that hangs over a G10 encoder plate on the rotor. The photodiodes are aligned with a matching set of five IR light-emitting diodes (LEDs) located below the plate. The encoder plate is slotted at two different radii in order to chop the light emitted by the LEDs. 
The photo current signals chopped by the 40 wider slots drive the motor coils by providing feedback through the motor drive electronics, while the signal chopped by the $570 - 1 = 569$ finer slots is fed to the encoder electronics for calculation of the rotation angle. One missing finer slot serves as a reference mark for the mechanical origin of the rotor.
A custom PCB driver board amplifies the photocurrent signal and generates electric signals for the encoder and motor coils. Additionally, it incorporates a frequency-to-voltage circuit to generate the rotation frequency feedback signal.
Two BeagleBone Black microcontrollers\footnote{BeagleBone Black, Beagle Board: \protect\url{https://beagleboard.org/black}} acquire the encoder data, one for each encoder assembly. It employs two programmable real-time units operating with an internal free-running 200\,MHz clock and records both the clock \change{counts} for the rising and falling edges of the amplified and digitized photo current signals as well as the observatory-wide master clock signal in IRIG-B format. This allows the assignment of precise master clock timestamps to each encoder sample. With typical rotation frequency of 2~Hz, the resulting data rate of the encoder \change{counts} is $2\times2\times570\simeq4$\,kHz, accounting for both rising and falling edges. 
The rotation frequency control is managed using a PID controller\footnote{Omega CNi16D54-EIT}, which receives a feedback signal from a driver board. 
A custom digital phase compensation unit is used to optimize drive efficiency.


\subsection{HWP rotation angle reconstruction method}
The procedure for estimating the rotation angle is divided into three steps: preprocessing, angle solution, and linear interpolation to detector timestamps. The science observations are conducted over a typical period of one hour, and we reconstruct the HWP rotation angle in the observation basis.

\subsubsection{Preprocessing}

\begin{figure*}
    \centering
    \begin{minipage}[h]{0.99\linewidth}
        \includegraphics[width = 0.99\textwidth]{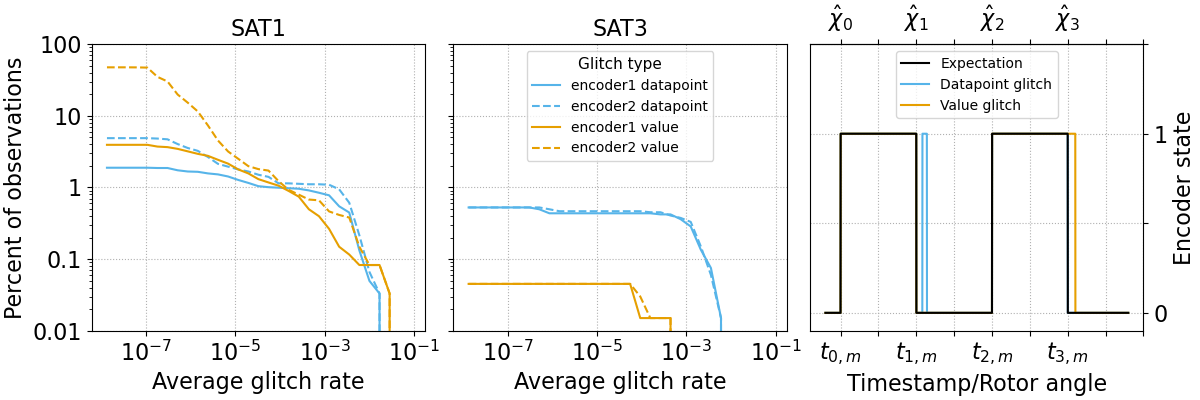}
    \end{minipage}
    \caption{The left and middle panels show the distribution of glitched encoder \change{counts} separated by glitch type and platform. A typical observation is one hour and contains approximately 10$^7$ encoder \change{counts}. Glitches found for SAT2 were minimal among three SATs, and not reported explicitly. 
    The right panel shows an idealized data point glitch and value glitch, along with an expected digitized encoder signal.}
    \label{fig:glitches}
\end{figure*} 

The first step is the preprocessing of the individual encoder \change{counts}. We assign a relative angle to each individual \change{count} within a single rotation. The edge of the reference slot is used to determine the angular offset with respect to the mechanical origin. Each \change{count} is then assigned a timestamp by referencing the data acquisition device's internal 200~MHz clock \change{counts} to the master clock signal. This creates an output of rotor angles $\hat{\chi}_i\equiv2\pi i/N$ with timestamps $t_{i,m}$, where $i$ is a time-ordered index, $N=1140$ is the total number of rising and falling edges of the encoder slots, and $m=0, 1$ are the labels of two encoders positioned 180$^\circ$ from one another. The index $i$ is defined so that $t_{i,0} \simeq t_{i,1}$, i.e. the 0-th timestamp corresponds to the 0-th ($N/2$-th) encoder slot for the encoder~0 (encoder~1).

The complications of preprocessing arise from the handling of non-idealities of data, i.e. removal and correction of glitches and filling of \change{counts} that are dropped during the network communication between the encoder and the data acquisition server. Glitches are defined as spurious \change{counts} or \change{counts} that deviate significantly from expected timing, and we classify these into ``data point glitches" and ``value glitches". 
Data point glitches often appear in pairs over short timescales and are believed to result from electrical noise or pickup in the raw encoder signal. This noise causes the signal to fluctuate around the digitization threshold, generating a false pair of \change{counts} for glitch's rising and falling edge\footnote{Occasionally, single data point glitch is observed. These lack a clear physical origin and are suspected to arise internally within the micro-controller used for data acquisition.}.
Value glitches can occur when a data point glitch overlaps with a real rising or falling edge, causing the recorded \change{count} of the rising (falling) edge to shift forward (backward) in time\footnote{Such a generation method is sufficient to explain the prevalence of value glitches in SAT2 and SAT3. For SAT1 the frequency of value glitches is larger, indicating an additional generation source, possibly a lower signal to noise ratio in the raw photo current signal or larger fluctuations in bias voltages supplied to the driver board.}. Figure~\ref{fig:glitches} shows the frequency of glitches for SAT1 and SAT3 along with an example illustration of how these glitches appear in the raw encoder \change{counts}. 
The incidence of these glitches is low: for 95\% of observations with SAT1 and over 99\% of observations with SAT2 and SAT3, the average glitch rate, defined as the fraction of glitches in the total number of data samples, is $\leq 10^{-6}$. 
Given the accuracy required for the rotation angle reconstruction, even such rare occurrences could be non-negligible and are carefully corrected, as described below.

Data point glitch corrections are applied to rotations that have more than $N$ \change{counts}, by removing \change{counts} that do not follow an expected pattern. 
Value glitches are corrected by comparing the pattern of \change{counts} for an individual rotation against a template constructed from an average across all rotations. Individual \change{counts} that differ from the template by more than 5\% of the expected incremental count are marked as value glitches and corrected\footnote{Vibration-induced variations that we describe in Sec.~\ref{sec:oc} are measured to be below the 5\% level.}. 

\subsubsection{Angle Solution} \label{sec:solution}

The second step is to solve for the rotation angle from the two sets of encoder timestamps, $t_{i,m}$. We denote the true angle of the rotor when $t_{i,m}$ is measured as $\chi_{i, m}$. The true time as a function of the true angle of the rotor, $\chi$, is $t(\chi)$. Then $t_{i,m}$ is expressed as
\begin{align}
    t_{i,m} &= t(\chi_{i, m}) + \mathcal{N} \nonumber \\
    &= t\left( \hat{\chi}_i - \theta_{i,m} - (-1)^m \phi_i \right) + \mathcal{N} \nonumber \\
    &\simeq t\left(\hat{\chi}_i\right) - \frac{dt}{d\chi} \theta_{i,m} - (-1)^m \frac{dt}{d\chi} \phi_i + \mathcal{N},
    \label{eq:timestamps_model}
\end{align}
where $\mathcal{N}$ is a noise, and $\theta_{i,m}$ is a periodic angle bias due to the deviation of the encoder slot pattern from the ideal, equally spaced pattern.
The $\theta_{i,m}$ shifts the two encoders' timestamps in the same way at different times, and the following relation holds because of its symmetry.\footnote{$\theta_{i,m}$ is non-zero because of the machining inaccuracies of the encoder plate and the non-ideal digitization threshold of the encoder signal. The non-ideal digitization threshold causes the rising edge and falling edges to shift in opposite directions by the same amount, but the amount of shift can be different between the two encoders. We take the average of two adjacent of rising and falling edges to cancel this bias. The template for $\theta_{i,m}$ is constructed after this pair averaging and thus is almost entirely defined by machining imperfections. We therefore assume the symmetry stated in Eq.~\eqref{eq:theta_symmetry}.} 
\begin{equation}\begin{aligned}\label{eq:theta_symmetry}
     \theta_{i,m} &= \theta_{i+N,m} && \text{for all $i$, $m$}, \\
     \theta_{i,0} &= \theta_{i+N/2,1} && \text{for all $i$}.
\end{aligned}\end{equation}
The $\phi_i$ term is an apparent angle shift due to the off-centering of the rotor perpendicular to the line connecting the two encoder heads. This effect causes the two encoders to have apparent offsets with opposite signs, as illustrated in Fig.~\ref{fig:method}(c). This effect should be removed because it does not rotate the HWP axis and does not contribute to the polarization modulation. The displacement of the rotor can be expressed as $R_\text{enc}\tan\phi_i \simeq R_\text{enc}\phi_i$, where $R_\text{enc}=346\,\text{mm}$ is the radius of the encoder slot (Fig.~\ref{fig:method} (b)). We use this to take precise measurements of the rotor position and its vibration (Sec.~\ref{sec:oc}).

The goal of the rotation angle solution is to estimate the true time when the angle of the rotor is $\hat{\chi}_i$:
\begin{align} \label{eq:t_hat}
    \hat{t}_i \simeq  t_{i,m} + \frac{dt}{d\chi} \theta_{i,m} + (-1)^m \frac{dt}{d\chi} \phi_i,
\end{align}
where $\hat{t}_i$ is the estimated true time.
To this end, we estimate unknown variables $\theta_{i, m}$ and $\phi_i$ from the two encoders' data as described in Appendix~\ref{app:solution}. 
The typical size of $\theta_{i, m}$ is $\leq0.1^\circ$, $\phi_i$ varies from $0^\circ$ to $0.7^\circ$ depending on the elevation of the telescope, and a single increment of the encoder angle is $2\pi/N \simeq 0.3^\circ$.

\subsubsection{Linear Interpolation}
The final step is to estimate the rotation angle at the 200\,Hz timestamps of the detector readout system, denoted as $t^\prime_j$. We linearly interpolate the encoder data, a set of $\hat{t}_i$ and $\hat{\chi}_i$, to $t^\prime_j$, and construct $\hat{\chi}(t^\prime_j)$. The final product is a set of $t^\prime_j$ and $\hat{\chi}(t^\prime_j)$.

Here we discuss potential biases in the reconstructed angle $\hat{\chi}(t^\prime_j)$. 
Interpolation to $t^\prime_j$ can potentially introduce a bias to the angle solution.
Focusing on a specific detector timestamp $t_0^\prime$, we consider the two adjacent angle encoder timestamps $t_{0,m}$ and $t_{1,m}$, which have a bias of $d_{0,m}$ and $d_{1,m}$, respectively. In addition, we consider a time offset between the HWP encoder and detector readout system $\Delta \equiv t^\prime-t$, \change{to account for possible effects such as miscalibration of the detector time constants or synchronization errors.} Then, $t_0'$ can be expressed as $t_0' = (1-s)t_{0,m} + st_{1,m} + \Delta$, where $s~(0\leq s \leq 1)$ is a parameter representing the relative distance of $t_{0,m}$ and $t_{1,m}$ to $t_0'$. The bias of the interpolated angle $\hat{\chi}(t^\prime_0)$ can be expressed, to the leading order of small parameters $d_{0,m}$ and $d_{1,m}$, as 
\begin{align}
    \delta\hat{\chi}(t^\prime_0) &\simeq -\frac{\hat{\chi}_1 - \hat{\chi}_0}{t_{0,m}-t_{1,m}} [(1-s)d_{0,m} + sd_{1,m} + \Delta] \nonumber \\
    &\simeq -\frac{d\chi}{dt}\big|_{t=t^\prime_0}[(1-s)d_{0,m} + sd_{1,m} + \Delta].
\end{align}
Substituting the timestamp bias caused by $\theta_{i,m}$ and $\phi_i$ (Eq.~\eqref{eq:timestamps_model}), the bias of $\hat{\chi}(t^\prime_0)$ without correction for either effect is
\begin{align}
    \delta\hat{\chi}(t^\prime_0)
    &\simeq (1-s)\theta_{0,m} + s\theta_{1,m} \nonumber \\
    &~~~~+ (-1)^m [(1-s)\phi_0 + s\phi_1] - \frac{d\chi}{dt}\Delta.
    \label{eq:angle_bias}
\end{align}
The angle biases caused by $\theta_{i,m}$ and $\phi_i$ do not depend on the rotation direction. On the other hand, the angle bias caused by $\Delta$ depends on the rotation direction and speed, $\frac{d\chi}{dt}$. 

\subsection{Half-wave plate polarimetry}
We describe the polarimetry using HWP polarization modulator.
The ideal continuously rotating HWP modulates the incident light with linear polarization as
\begin{align}
    d_{m} = I + \mathrm{Re} [\left(Q + iU\right)m(\chi)],
\end{align}
where $I$, $Q$, and $U$ are the Stokes parameters of the incident light, $\chi(t)$ is the rotation angle of the HWP with typical rotation frequency of 2~Hz, and $m(\chi) = \exp(-i4\chi)$ is the modulation function. To reconstruct the input $Q$ and $U$, we band-pass filter the modulated data with a typical bandwidth of two to four times the HWP rotational frequency around the modulation frequency of 8~Hz, multiply by the demodulation function, $2m(-\hat{\chi})$, constructed with the measured angle $\hat{\chi}$, and apply a low-pass filter with a typical cutoff frequency at the HWP rotation frequency or its double. We call this procedure demodulation \citep{kusaka_modulation_2014}.

\subsection{Requirement of HWP angle reconstruction} \label{sec:requirement}
This section discusses the requirement of the reconstructed HWP rotation angle. 
A bias $\delta\hat{\chi} \equiv \hat{\chi} - \chi$ in the reconstructed angle shifts the polarization from the ideal $Q+iU$ to $(Q+iU)\exp(i4\delta\hat{\chi})$, corresponding to a rotation of the polarization angle by $2\delta\hat{\chi}$. We decompose $\delta\hat{\chi}$ into a constant component and a varying component. 
\change{The constant component of $\delta\hat{\chi}$ is the offset between the encoder origin and the HWP fast axis. The constant component, together with the detector's intrinsic polarization angle, determines the SAT's absolute polarization angle calibration.  Since these two quantities are degenerate, the constant component itself is not required to be known precisely, nor can it be calibrated well from observational data; it is known only to an accuracy of a few degrees from laboratory measurements and fabrication notes. On the other hand, the absolute polarization angle is calibrated using, for example, a sparse wire grid calibrator, a drone-based calibrator, the Crab Nebula, or self-calibration with CMB data~\citep{Nakata_2026, Coppi_2025, Aumont2020, Keating_2013}, }and its calibration requirement is approximately $0.1^\circ$ \citep{10.1117/12.2313832}. 

The varying component of $\delta\hat{\chi}$ is required to satisfy $\sigma_\chi < 3\,\mu$rad$\sqrt{\text{s}}$ \citep{SO_CHWP}. This requirement is derived from the noise injected into the demodulated detector data by $\delta\hat{\chi}$, given by $2\sqrt{2} A_4\sigma_\chi$ \citep{HillPB2bCHWP2020}.  We require this noise to be an order of magnitude lower than the instantaneous sensitivity of a single SAT with MF bands, combining both 90 and 150~GHz bands conservatively. 
Here $\sigma_\chi$ is the noise density of $\delta\hat{\chi}$ and $A_4$ is the amplitude of instrument polarization, primarily caused by optical components skyward of the HWP. This noise is caused by the small mis-demodulation of $A_4$ due to $\delta\hat{\chi}$, and is largely common between detectors and therefore does not average down between detectors.
The impact of the varying components of $\delta\hat{\chi}$ is generally smaller than that of the constant component as they average down in celestial coordinates. In particular the components with frequency larger than half of the HWP rotation frequency are strongly suppressed, as they are outside of the demodulation bandwidth. 

\section{Results} \label{sec:results}
This section describes the rotation performance of the three currently deployed HWP polarization modulators during the initial year of observations. Specifically, we describe the angle reconstruction accuracy, stability, optical validation of the off-centering correction, and vibrations.

\subsection{Encoding accuracy} \label{sec:accuracy}
\begin{figure*}
    \centering
    \begin{minipage}[h]{0.49\linewidth}
        \includegraphics[width = 0.99\textwidth]{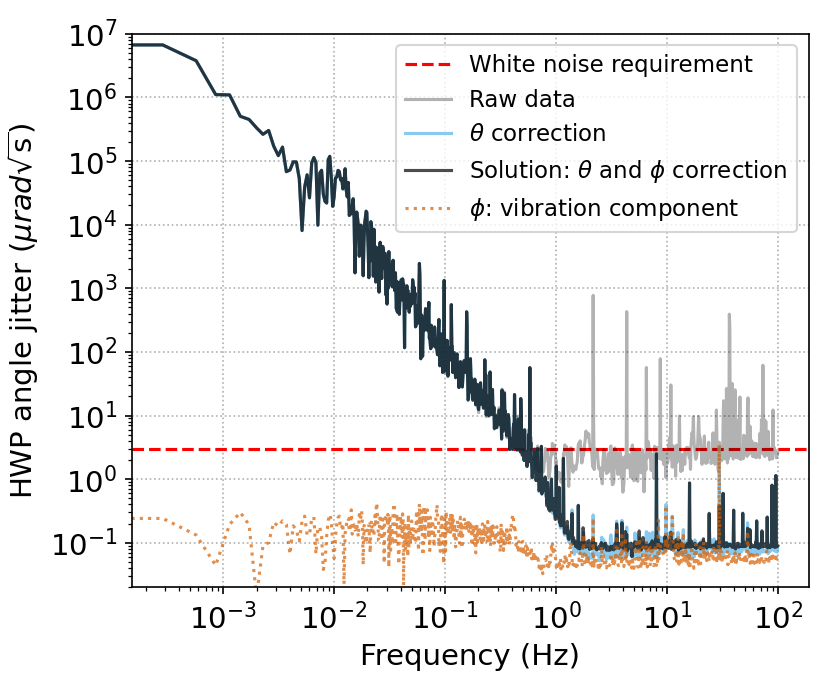}
    \end{minipage}
    \begin{minipage}[h]{0.49\linewidth}
        \includegraphics[width = 0.99\textwidth]{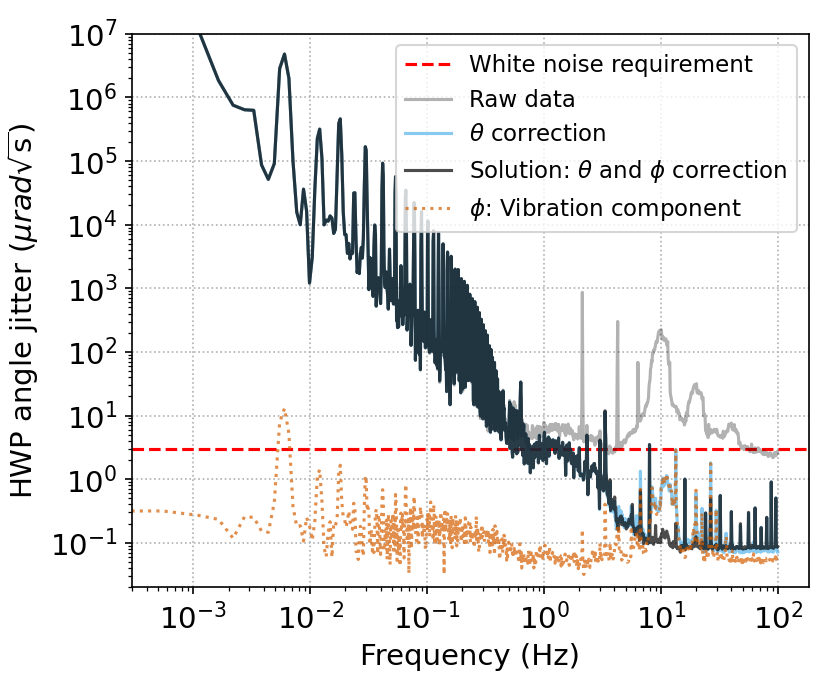}
    \end{minipage}
    \caption{Amplitude spectral density of the rotation angle jitter for a 2\,Hz rotation, measured over approximately one hour of data for SAT3. Left panel: The spectrum when the telescope is stationary. The gray line shows the spectrum of raw data. The $1/f$ noise is due to rotation speed variation and is not an encoding error. The peak at 2\,Hz and its harmonics along with the elevated noise above 2\,Hz are due to the periodic bias of the rotation angle. The blue and black lines, representing the spectrum with corrections, exhibit a flat white noise profile. Right panel: The spectrum when the telescope is scanning. The scan-induced modulation of the rotation seen at 6\,mHz and its harmonics are not encoding errors. 
    The scan-induced vibrations of the rotor at around 10\,Hz is seen in blue line, and it is successfully subtracted in the black line representing the spectrum of the solution. The orange line shows the effect of vibration subtracted from the blue line, which corresponds to the difference of angle measured by two encoders (Eq.~\eqref{eq:vibration_comp}). The horizontal red dashed line marks the requirement of noise level.}
    \label{fig:spectra}
\end{figure*} 

Figure~\ref{fig:spectra} shows the amplitude spectral density of the rotation angle jitter for a 2\,Hz rotation, measured over approximately one hour. The jitter is defined as the deviation from smooth rotation, $\hat{\chi}(t^\prime) - \overline{\frac{d\hat{\chi}}{dt^\prime}}t^\prime$.
The left panel shows the spectrum when the telescope is stationary. The $1/f$ noise is due to the rotation speed variation that we discuss in Sec.~\ref{sec:stability} and is not an encoding error. The peak at 2\,Hz and its harmonics along with the elevated noise above 2\,Hz are caused by the periodic bias of the rotation angle. After their correction, the spectrum exhibits a flat white noise profile, which validates the correction processes. The right panel of Fig.~\ref{fig:spectra} shows the spectrum when the telescope is scanning. The peak at 6\,mHz and its harmonics are due to the scan-induced modulation of the rotation speed which we discuss in Sec.~\ref{sec:stability}. These are not encoding errors. The broad peak around 10\,Hz is due to the scan-induced vibrations of the rotor which we discuss in Sec.~\ref{sec:oc}. This effect results in apparent changes of rotation angle, and the flat white noise profile after its subtraction validates the correction process. We describe the detailed analysis in Appendix~\ref{app:solution}.

Figure~\ref{fig:hwp_angle_white_noise} shows the cumulative distribution of encoder angle noise, $\sigma_\chi$. It is estimated for every observation, averaging the spectrum between 10 and 100\,Hz including residual peaks. The median noise level is 0.16\,$\mu$rad$\sqrt{\text{s}}$ and we achieved the requirement of 3\,$\mu$rad$\sqrt{\text{s}}$ in more than 99.9\% of observations. 

\begin{figure}
    \centering
    \includegraphics[width = 0.48\textwidth]{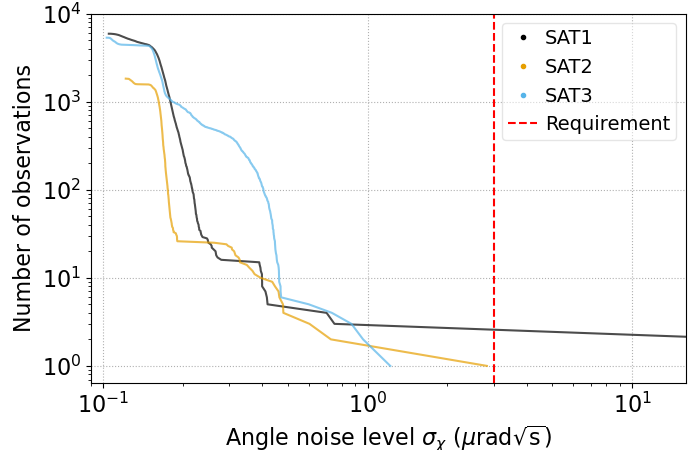}
    \caption{Cumulative distribution of reconstructed angle noise level per observation. The noise is estimated by averaging the noise spectral density between 10 and 100\,Hz. The vertical red dashed line shows the required noise level. More than 99.9\% of observations satisfy the requirement. The median noise level is 0.16\,$\mu$rad$\sqrt{\text{s}}$.}
    \label{fig:hwp_angle_white_noise}
\end{figure}

\subsection{Stability} \label{sec:stability}
This section provides a detailed analysis of the rotational speed stability performance at 2\,Hz, and its possible future improvements.
The performance can be naturally assessed on three relevant timescales associated with the operation of the system and data acquisition: a single revolution of 0.5~second, a typical observation of one hour, and longer time scales beyond a single observation. 
Rotational stability on different timescales affects our data and analysis strategy in distinct ways. 

The rotational speed variation within a single revolution is primarily due to the deviation of the center of mass of the rotor from its geometric center. This is periodic over an observation, so it is subtracted in a same way as the encoder slot non-uniformity as we describe in more detail in Appendix~\ref{app:solution}. The subtraction of this speed variation leads to a $<0.1^\circ$ level of bias, and its impact is negligible because as it is outside of the bandwidth of demodulation, as discussed in Sec.~\ref{sec:requirement}.

\begin{figure}
    \centering
    \includegraphics[width = 0.48\textwidth]{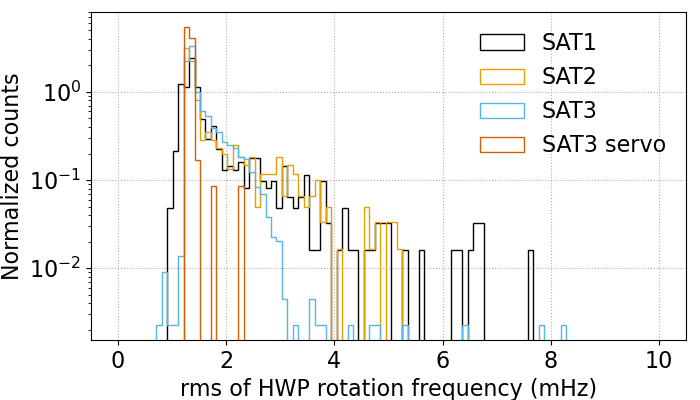}
    \caption{Normalized histogram of the rms of the rotation frequency per observation in the initial year of observations. Observations are conducted at 60$^\circ$ elevation and and a scan speed of 0.5\,$^\circ$/s for approximately one hour. The maximum of the distribution at 1.3\,mHz is due to the scan-induced variation.}
    \label{fig:rms_fhwp}
\end{figure}

On the time scale of a typical observation of one hour, the rotation speed modulation due to telescope scanning is the dominant effect on stability. Figure~\ref{fig:rms_fhwp} shows a normalized histogram of the rotation frequency stability, quantified as the rms variation within each observation. Observations are conducted at 60$^\circ$ elevation and and a scan speed of 0.5\,$^\circ$/s. 
The maximum of the distribution at 1.3\,mHz is caused by scan-induced rotational speed modulations modeled in \citet{SO_CHWP} as $ \frac{d\chi}{dt} = \overline{\frac{d\chi}{dt}} - \frac{d\theta_\text{az}}{dt}\sin\theta_\text{el}$,
where $\overline{\frac{d\chi}{dt}}$ is the average angular speed of the rotor, and $\theta_\text{az}$ and $\theta_\text{el}$ are the azimuth and elevation of the telescope\footnote{The $\sim$1.3~mHz corresponds to the scan speed projected along the rotor spin axis in Hz.}.
The achieved stability of 1.3~mHz is comparable to the frequency resolution over an observation and is better than the typical $1/f$ knee frequency of the detector data. This level of performance helps to simplify data analysis processes. In principle, variations in rotational speed do not pose an issue in analysis as long as the rotation angle reconstruction is accurate.

\begin{figure}
    \centering
    \includegraphics[width = 0.48\textwidth]{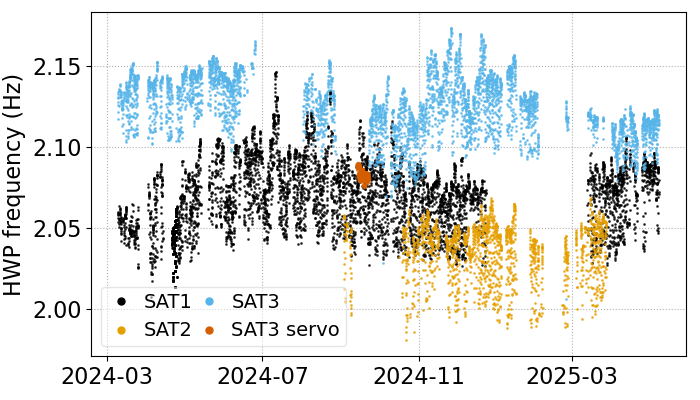}
    \caption{Rotation frequencies in the initial year of observations. The red points represent data taken when we placed a servo heater in the enclosure of the control electronics to eliminate the temperature dependent component of the rotation variation.}
    \label{fig:stability_all}
\end{figure}

\begin{figure}
    \centering
    \includegraphics[width = 0.48\textwidth]{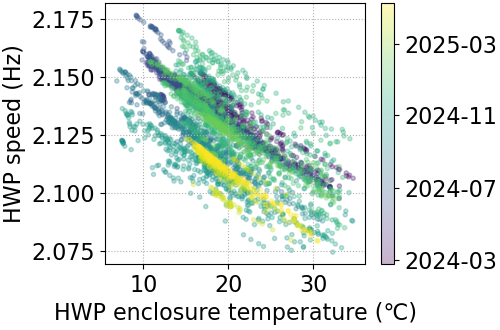}
    \caption{Rotation frequencies of SAT3 rotor as a function of local time. Clear correlation is seen between the rotation speed and the temperature of the control electronics enclosure with correlation coefficient of $-2$\,mHz/$^\circ$C.}
    \label{fig:temperature_dependence}
\end{figure}

\begin{figure}
    \centering
    \includegraphics[width = 0.48\textwidth]{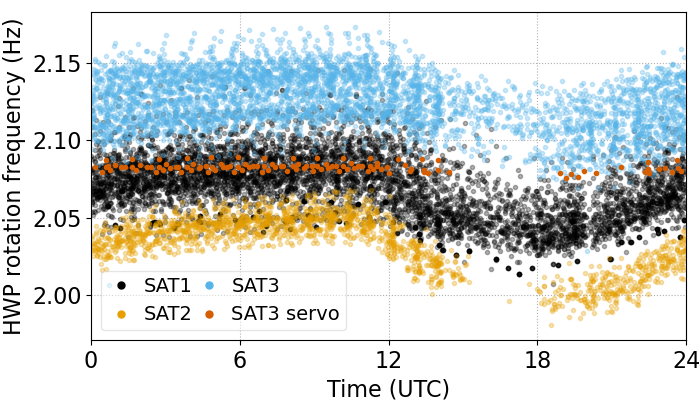}
    \caption{Rotation frequencies as a function of local time. The red points represent data taken when we placed a servo heater in the enclosure of the control electronics to eliminate the temperature dependent component of the rotation variation.}
    \label{fig:stability_diurnal}
\end{figure}

Figure~\ref{fig:stability_all} shows the rotation frequency over a one-year period. It has a variation of approximately 100\,mHz around 2\,Hz. 
While the original requirement on the long time scale stability was 10\,mHz due to concerns about avoiding mechanical resonances \citep{SO_CHWP},
we have found that this was too conservative and that a year-long stability of 100\,mHz is sufficient for our needs.
This longer time scale variation is partially attributed to the temperature dependence of the control electronics. Figure~\ref{fig:temperature_dependence} illustrates a significant correlation between temperature variations inside the enclosure of the control electronics and the rotation frequency, with a correlation coefficient of $-2$\,mHz/$^\circ$C. The temperature dependence primarily induces the diurnal variation of the rotation frequency as shown in Fig.~\ref{fig:stability_diurnal}. The temperature dependence of the rotation frequency is additionally confirmed by installing a servo heater within the enclosure for a week to maintain the temperature at the daily maximum; these data points are shown in red in Fig.~\ref{fig:stability_all} and Fig.~\ref{fig:stability_diurnal}. 
However, the temperature changes are not sufficient to fully explain all variations in the rotation frequency. A possible future improvement of long time scale stability would be altering the control scheme to use the rotation frequency measured by the encoder as a feedback signal. This change would ensure that the feedback signal is identical to the rotation frequency measured by the observatory wide master clock and would eliminate frequency variations.

\subsection{Off-center displacement} \label{sec:oc}
The off-center displacement of the rotor is divided into a constant component and a varying component. The constant component is caused by gravity and depends on the elevation of the telescope. The varying component is caused by the vibration of the telescope. 
The component of the displacement perpendicular to the line connecting the two encoders induces an apparent offset, $\phi_i$, to the measured angle, as discussed in Sec.~\ref{sec:solution}. Telescope boresight angle rotation allows us to measure the orthogonal component of the displacement. The precision of the displacement measurement is $R_\text{enc}\sigma_\chi=0.06\,\mu\text{m}\sqrt{\text{s}}$.

\begin{figure}
    \centering
    \includegraphics[width = 0.49\textwidth]{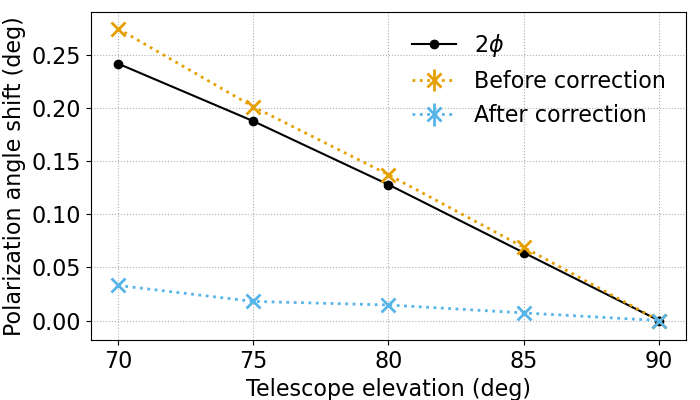}
    \caption{The measured polarization angle of the fixed sparse wire grid calibrator as a function of telescope elevation with and without correction of $\phi_i$. The solid black line shows the expected shift of polarization angle due to $\phi_i$. Without correction, the measured polarization angle rotates depending on elevation. The measured polarization angle after correction is almost constant with elevation. This measurement was conducted with SAT3.}
    \label{fig:optical_sag_evaluation}
\end{figure}

The correction of the component constant during an observation is particularly important as it could induce an elevation dependent polarization angle bias. This must be corrected within the SAT's absolute polarization angle requirement of approximately $0.1^\circ$ \citep{10.1117/12.2313832}. The statistical uncertainties of this correction are negligible, as $\sigma_\chi/\sqrt{T} \ll 0.1^\circ$, where $T$ is the duration of the observation.
To optically verify this correction, we placed a sparse wire grid calibrator with a fixed angle on top of the window to inject a constant linear polarization \citep{Sparse_WG, Nakata_2026}. We then changed the elevation of the telescope from 90$^\circ$ to 70$^\circ$ by 5$^\circ$ steps to induce the change in $\phi_i$. The solid black line of Fig.~\ref{fig:optical_sag_evaluation} shows the expected shift of polarization angle due to $\phi_i$. The rotor displaced by approximately 0.73\,mm from elevation 90$^\circ$ to 70$^\circ$, which corresponds to $\phi_i=0.24^\circ$. 
The yellow and blue lines in Fig.~\ref{fig:optical_sag_evaluation} show the change of the measured polarization angle with respect to the telescope elevation, with and without correction for $\phi_i$.
As expected, the measured polarization angle shifts depending on elevation. After correction, the change of the measured polarization angle is smaller than the requirement. The residual polarization angle shift, smaller than $0.05^\circ$, may be attributed to the increase of detector time constants, intensity to polarization leakage, and a sag of wire grid, all of which depend on the elevation. The residuals are conservative estimates of the correction errors as the SATs perform constant elevation scans for science observations and calibrate detector time constants at that fixed elevation. 

\begin{figure}
    \centering
    \includegraphics[width = 0.49\textwidth]{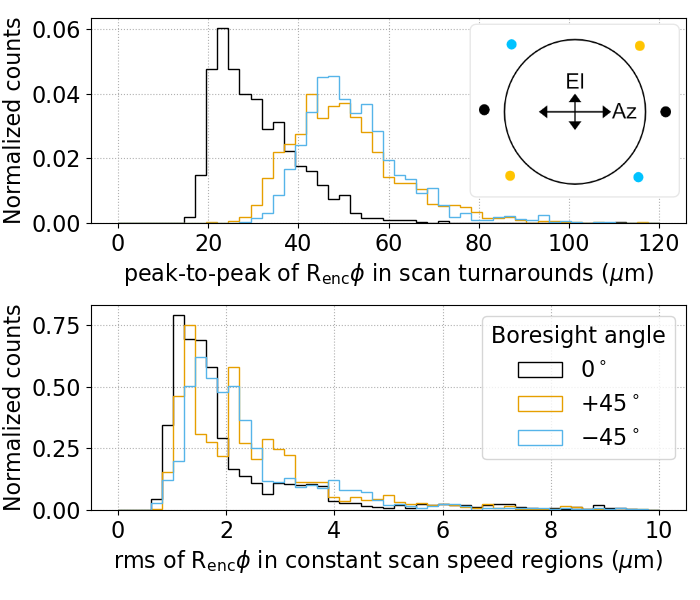}
    \caption{Normalized histogram of the vibration-induced displacement of the SAT1 rotor, perpendicular to the line connecting the two encoders, during different phases of the constant elevation scans, quantified as $R_\text{enc}\phi_i$ (Sec.~\ref{sec:solution}). Top panel: Peak-to-peak of the rotor displacement at the scan turnarounds. The measured vibration-induced displacement is larger when the boresight angle is nonzero. Top right panel: Schematic diagram of the encoder head locations relative to the rotor at boresight angles of $0^\circ$ and $\pm45^\circ$. The encoder heads and rotor are indicated by small circles and a large black circle, respectively. The displacement is larger in the azimuth (scan) direction regardless of the boresight angle. When the boresight angle is zero, $\phi_i$ is only sensitive to the displacement in the elevation direction, whereas at $\pm45^\circ$, $\phi_i$ is equally sensitive to displacements in both azimuth and elevation directions, resulting in larger measured displacements. Bottom panel: Root mean square of vibration-induced displacement in constant scan speed regions.}
    \label{fig:vib_hist}
\end{figure}

The motions of the telescope excite vibrations of the rotor at the characteristic vibration frequencies of its superconducting magnetic bearing. The characteristic frequencies parallel and perpendicular to the optical axis of the SAT are 9\,Hz and 13\,Hz, respectively \citep{SO_CHWP}. 
Figure~\ref{fig:vib_hist} shows the amplitude of the vibration-induced displacement of the rotor, perpendicular to the line connecting the two encoders, during the constant elevation scans, quantified as $R_\text{enc}\phi_i$ (Sec.~\ref{sec:solution}). The   
vibration-induced displacement at the scan turnarounds are below 100\,$\mu$m peak-to-peak, and typical vibration-induced displacement at constant scan speed regions are 1\,$\mu$m rms. The orange line in Fig.~\ref{fig:spectra} shows the spectrum of these vibrations, which are successfully subtracted as can be seen from the black line in the same figure. 

While these vibrations do not contribute to polarization modulation, they are around the modulation frequency of 8\,Hz. This generates the potential for vibrational coupling or magnetic pickup in the detectors and contaminate the demodulated polarization data. This effect is expected to average down because the phases of the vibrations and the rotations are random, but the actual effects remain to be investigated. Our analysis is also limited to two-dimensional measurements of the rotor vibrations. Additional possible mechanisms of the rotor vibrations are discussed in Appendix~\ref{app:forces}.

\section{Conclusion} \label{sec:conclusion}
This study advances the rotation angle reconstruction method for the Simons Observatory SAT HWP polarization modulators. We demonstrate a detailed evaluation of the HWP rotation performance, especially regarding its angle reconstruction accuracy, its stability and its vibrations. We achieve required angle accuracy in more than 99.9\% of observations, demonstrating the robustness of all three deployed systems. This is a crucial step towards achieving the SAT science goals. Based on the accuracy of the HWP angle solution presented here, we do not need to account for its uncertainty in the analysis of CMB polarization corresponding to the $r\simeq$0.001 level.

This detailed performance evaluation reveals several avenues for future improvements to the polarization modulator systems. Enhancements to the encoder circuitry to minimize glitches that contribute to angular noise would be desirable. We have outlined the path for improving stability by identifying the causes of instability. Furthermore, the vibration analysis indicates that designing the SMB to ensure that its characteristic vibration frequency is sufficiently above the modulation frequency would mitigate potential systematic errors, although a detailed study of these effects is left to future investigations. Additionally, analysis of periodic rotational fluctuations demonstrates the potential to characterize nonuniformities in the SMB and its rotation mechanism, suggesting design enhancements that enable more accurate assessments of these nonuniformities.


\begin{acknowledgments}
This work was supported by MEXT KAKENHI Grant Numbers JP19H00674, JP23H00105, JP23H01202, JP22H04913, and by JSPS KAKENHI Grant Numbers JP24K23938, JP24KJ0663, JP25KJ0840, and by the JSPS Core-to-Core Program JPJSCCA20200003. 
This work was supported by World Premier International Research Center Initiative (WPI).
This work was supported in part by a grant from the Simons Foundation (Award \#457687, B.K.). 
Work at Princeton was supported by the David Wilkinson Research Fund.
Work at LBNL was supported in part by the U.S. Department of Energy, Office of Science, Office of High Energy Physics, under contract No. DE-AC02-05CH11231.
The work reported on in this paper was performed using Princeton University’s Research Computing resources and resources of the National Energy Research Scientific Computing Center (NERSC). J.S. acknowledges the support from the International Graduate Program for Excellence in Earth-Space Science (IGPEES) and the JSR Fellowship, the University of Tokyo. D.S. acknowledges the support from FoPM, WINGS Program and the JSR Fellowship, the University of Tokyo. 
We thank Jansen Ball for improving the motor drive electronics. 

\end{acknowledgments}

\appendix
\section{Solution} \label{app:solution}
This section describes the details of the rotation angle reconstruction method, specifically the estimation $\hat{t}_i$ of Eq.~\eqref{eq:t_hat}. The first step is to estimate $\theta_{i, m}$ and correct the biases due to $\theta_{i, m}$.
We approximate the smoothed timestamp $\bar{t}_i$ by the cubic spline interpolation on the sub-sampled timestamps $t_{i,m}$ at indices $i=0,~N,~2N...$, followed by averaging the resulting interpolated values from the two encoders. The deviation of $t_{i,m}$ from $\bar{t}_i$ is expressed as 
\begin{align}
    T_{i,m} &\equiv t_{i,m} - \bar{t}_i \simeq \Delta t_i - \frac{d\bar{t}_i}{d\chi} \theta_{i, m} - (-1)^m \frac{d\bar{t}_i}{d\chi} \phi_i,
\end{align}
where $\Delta t_i$ is a fluctuation of the rotation within a revolution. To estimate periodic bias $\theta_{i,m}$, we multiply $T_{i,m}$ by the rotation speed approximated by the inverse of cubic spline derivative of $\bar{t}_i$ and taking average of every $N$-th points as 
\begin{align}
    \Psi_{j,m} &\equiv \left\langle \frac{d\chi}{d\bar{t}_i} T_{i,m} \right\rangle_{j,N} \nonumber \\
    &= \left\langle \frac{d\chi}{d\bar{t}_i} \Delta t_i \right\rangle_{j,N}
    - \left\langle \theta_{i, m} \right\rangle_{j,N}
    - (-1)^m \left\langle \phi_i \right\rangle_{j,N} \nonumber \\
    &= \left\langle \frac{d\chi}{d\bar{t}_i} \Delta t_i \right\rangle_{j,N}
    - \theta_{j, m}
    - (-1)^m \left\langle \phi_i \right\rangle_{j,N} \label{eq:fluct}
\end{align}
where $j$ is an index from $0$ to $N-1$ and $\langle\cdot\rangle_{j,N}$ is an average at indices $i=j,~j+N,~j+2N...$. Due to the symmetry of $\theta_{i, m}$ given by Eq.~\eqref{eq:theta_symmetry}, the relation $\theta_{j, m} = \left\langle \theta_{i, m} \right\rangle_{j,N}$ holds. 
The $\Psi_{j,m}$ is biased from $\theta_{j, m}$ due to the periodic components of $\phi_i$ and $\Delta t_i$, but a bias due to $\phi_i$ is harmless, as it accounts for the effects of the periodic component of $\phi_i$. In contrast, a bias due to $\Delta t_i$ is formally undesirable as this is an actual fluctuation of rotation within a revolution. 
However, its impact is negligible because this is outside of the bandwidth of demodulation as discussed in Sec.~\ref{sec:requirement}, and also its amplitude is smaller than 0.1$^\circ$ as we show in Appendix~\ref{app:fluct}. 
Therefore, in practice, we construct the timestamps with correction of $\theta_{i,m}$ as 
\begin{align}
    \tilde{t}_{i,m} = t_{i,m} - \frac{d\bar{t}_i}{d\chi} \Psi_{i\,(\text{mod}\,N),m}.
\end{align}
We discuss the separation of components of $\Psi_{j,m}$ and the investigation of their origin in Appendix~\ref{app:fluct}. 

Finally, we remove the biases caused by non-periodic component of $\phi_i$, the non-periodic offcentering of the rotor caused by the vibrations, by taking the average of timestamps between two encoders.
The final solution of timestamps $\hat{t}_{i}$ is 
\begin{align}
    \hat{t}_{i} = \left(\tilde{t}_{i,0} + \tilde{t}_{i,1}\right) / 2.
\end{align}
The bias caused by vibration of the rotor is expressed as 
\begin{align} \label{eq:vibration_comp}
    \frac{dt}{d\chi}\hat{\phi_i} = \left(\tilde{t}_{i,0} - \tilde{t}_{i,1}\right) / 2.
\end{align}

\section{Analysis of periodic rotation fluctuation} \label{app:fluct}
This section describes the analysis of the components of the periodic bias of the timestamps $\Psi_{j,m}$ in Appendix~\ref{app:solution}. For the science analysis, $\Psi_{j,m}$ did not need to be separated into components but it is worth investigating further to provide feedback for the future development of the superconducting bearings and their rotation mechanisms.

To separate three components of $\Psi_{j,m}$, we consider decomposing $\Psi_{j, m}$ into common and differential mode of two encoders as $\Psi_{j,0} = (C_j + D_j) / 2$, $\Psi_{j,1} = (C_j - D_j) / 2$. From Eq.~\eqref{eq:fluct}, $C_j$ and  $D_j$ are expressed as 
\begin{align}
    C_j &= \left\langle \frac{d\chi}{d\bar{t}_i} \Delta t_i \right\rangle_{j,N} 
    - \left( \theta_{j, 0} + \theta_{j, 1} \right)/2, \\
    D_j &=  
    - \left( \theta_{j, 0} - \theta_{j, 1} \right)/2 - \left\langle \phi_i \right\rangle_{j,N}.
\end{align}
Then, we consider the normalized discrete Fourier amplitudes of these values.
We denote the normalized discrete Fourier amplitudes of $C_j$, $D_j$, $\left\langle \frac{d\chi}{d\bar{t}_i} \Delta t_i \right\rangle_{j,N}$, $\theta_{j, m}$ and $\left\langle \phi_i \right\rangle_{j,N}$ as $\tilde{C}_k$, $\tilde{D}_k$, $\tilde{T}_k$, $\tilde{\Theta}_{k,m}$ and $\tilde{\Phi}_k$, respectively\footnote{For example, \change{$\tilde{C}_k \equiv\frac{2}{N}\sum_{j=0}^N C_j \exp(-2\pi ijk/N)$.}}.
Due to the symmetry of $\theta_{i, m}$ given by Eq.~\eqref{eq:theta_symmetry}, the relation $\tilde{\Theta}_{k,0} = (-1)^k \tilde{\Theta}_{k,1}$ holds.
Therefore, we can classify the odd and even $k$-modes of $\tilde{C}_k$ and $\tilde{D}_k$ as follows:
\begin{equation}\begin{aligned}
    \tilde{C}_k &= \tilde{T}_k                              &&\text{if $k$ is odd} \\
    \tilde{C}_k &= \tilde{T}_k - \tilde{\Theta}_{k,0}       &&\text{if $k$ is even} \\
    \tilde{D}_k &= -\tilde{\Theta}_{k,0} - \tilde{\Phi}_k   &&\text{if $k$ is odd} \\
    \tilde{D}_k &= -\tilde{\Phi}_k                          &&\text{if $k$ is even}
\end{aligned}\end{equation}
The contribution of two individual components cannot be uniquely determined for even $k$-modes of $\tilde{C}_k$ and odd $k$-modes of $\tilde{D}_k$. However, $\tilde{T}_k$ and $\tilde{\Phi}_k$ are dynamical quantities that are expected to depend on parameters such as the elevation of the telescope and the rotation speed of the rotor. In contrast, $\tilde{\Theta}_{k,0}$ should be constant, independent of these parameters. Therefore, the dependence on elevation or rotation speed can be used to distinguish between these components.

\begin{figure*}
    \centering
    \begin{minipage}[h]{0.49\linewidth}
        \centering
        \includegraphics[width = 0.99\textwidth]{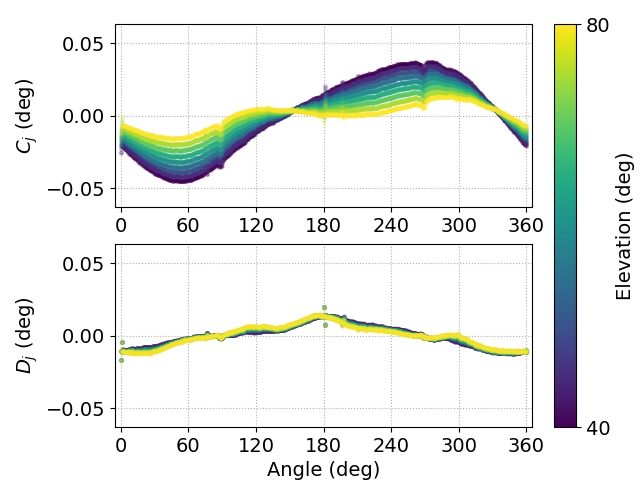}
    \end{minipage}
    \begin{minipage}[h]{0.49\linewidth}
        \centering
        \includegraphics[width = 0.99\textwidth]{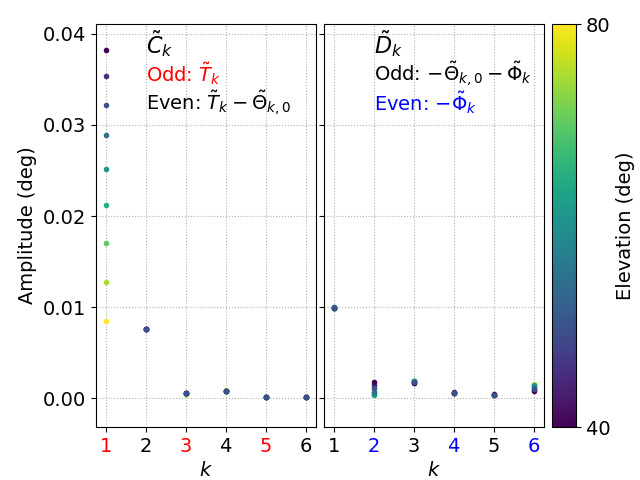}
    \end{minipage}
    \caption{Periodic rotation fluctuation of the SAT2 rotation mechanism. Left panel: Telescope's elevation dependence of $C_j$ and $D_j$. Right panel: Telescope's elevation dependence of lowest six discrete Fourier modes' amplitudes.}
    \label{fig:fluct}
\end{figure*}

Figure~\ref{fig:fluct} shows $C_j$ and $D_j$ and its lowest six $k$-modes of $\tilde{C}_k$ and $\tilde{D}_k$ and their elevation dependence. The primary mode that exhibits elevation dependence is $\tilde{C}_1$. This is angle-dependent rotation fluctuation due to the displacement of the rotor's center of mass from its geometric center.
Assuming only $\tilde{T}_1$, the $\chi$ can be written as
\begin{align}
    \chi = \overline{\frac{d\chi}{dt}} t + \chi_0 + |\tilde{T}_1| \sin \left(\overline{\frac{d\chi}{dt}} t + \chi_0\right)
\end{align}
where $\chi_0$ is the angle of the center of mass from the rotor's geometric center.
It thus follows
\begin{align}
    \frac{d\chi}{dt} \simeq \overline{\frac{d\chi}{dt}} (1+|\tilde{T}_1| \cos \chi).
\end{align}
We consider the case where the energy input from the motor and the energy loss from the friction are balanced. We denote the distance between the rotor's center of mass and its geometric center as $r_\text{cm}$. From the conservation law of energy, the following holds.
\begin{align}
    &\frac{I}{2}\left\{\overline{\frac{d\chi}{dt}}(1+ |\tilde{T}_1|\cos\chi) \right\}^2 \nonumber \\
    &= \frac{I}{2} \left(\overline{\frac{d\chi}{dt}}\right)^2 + mgr_\text{cm}\cos\theta_\text{el}\cos\chi,
     \label{eq:center_of_mass}
\end{align}
where $I$ is the momentum of inertia, $m$ is the mass of the rotor and $g$ is the acceleration of gravity. $I$ is well approximated by $I=mr^2_{I}/2$, where $r_I\simeq0.25$\,m is the effective radius of the rotor. Since $\tilde{C}_1 = \tilde{T}_1$, the $r_\text{cm}$ is expressed as 
\begin{align}
    r_\text{cm} = r_I^2\left(\overline{\frac{d\chi}{dt}}\right)^2|\tilde{C}_1|/2g\cos\theta_\text{el}.
    \label{eq:r_cm}
\end{align}
The $r_\text{cm}$ estimated by Eq.~\eqref{eq:r_cm} varies from 0.1\,mm to 0.4\,mm depending on the systems.
The rotation of the center of mass with radius $r_\text{cm}$ drives periodic force to the rotor and contributes to the $\tilde{\Phi}_k$.

Another benefit of analyzing $\tilde{C}_k$ is that it provides a precise probe for the non-uniformity of the SMB and its rotation mechanism. We consider the angle-dependent kinetic potential as $V(\chi)$, and its normalized discrete Fourier amplitudes as $\tilde{V}_k$.
As a generalization of Eq.~\eqref{eq:center_of_mass}, the following relation holds
\begin{align}
    \tilde{V}_k = I \left(\overline{\frac{d\chi}{dt}}\right)^2 \tilde{T}_k.
\end{align}
Therefore, in order to measure $\tilde{V}_k$, slow rotation is helpful to increase the observable $\tilde{T}_k$. 

The origins of $\tilde{V}_k$ are the non-uniformity of the SMB and the rotation mechanism. For the systems of SATs, the non-uniformity of the magnetic fields of the SMB can make the 32nd $k$-mode of $\tilde{V}_k$, motor drive sprocket magnets and motor drive coils can make 40th or 120th $k$-modes of $\tilde{V}_k$. To avoid ambiguity with the non-uniformity of encoder slots and better characterize non-uniformity of SMB and its rotation mechanism, it would be desirable to make the number of mechanical components and number of encoders relatively prime. 


\section{Forces on rotor} \label{app:forces}
This section enumerates the various forces that act on the rotor. While not all of these forces have been verified through measurements, their consideration is worthwhile, as any force acting on the rotor impacts its rotational performance or contributes to biases in rotation angle measurements. In addition, this study is important to ensure the instrument safety of the polarization modulator.

Figure~\ref{fig:forces} shows the schematic figure of the various forces acting on the rotor. The largest force is gravity. This causes elevation-dependent off-centering of the rotor as discussed in Sec.~\ref{sec:oc}. The rest of the forces are induced by the scans of the telescope. During constant-speed scans, the rotor is subject to a centrifugal force. Because the rotor is rotating, a gyroscopic force is also present, leading to a scan-direction-dependent tilt of the rotor. At turnarounds, the azimuthal acceleration of the telescope appears, while the centrifugal force and the gyroscopic force are temporarily absent, both effects contributing to vibrations of the rotor.

\subsection*{Gravity}
With the rotor mass \(m\) and gravitational acceleration \(g\), a force of \(\bm{F}_\text{grav}=-mg \hat{z}\) acts vertically downward.
In the absence of telescope scans, this force is exactly balanced by the supporting force of the bearing.

\subsection*{Centripetal force} 
During a constant speed scan at angular velocity \(\Omega\), gravity and the bearing force are not perfectly balanced, resulting in a net force directed from the rotor toward the scan axis. This net force serves as the centripetal force that governs the center-of-mass motion of the rotor. Its magnitude is
\begin{align}
    \bm{F}_\text{cent} = m D\Omega^2\cos\theta_\text{el}\,\hat{x}.
\end{align}
In the rotating frame of the telescope, this can equivalently be interpreted as a centrifugal force acting on the rotor. The magnitude of \(\bm{F}_\text{cent}\) is approximately 0.02\% of \(\bm{F}_\text{grav}\), with representative parameters of \(\Omega=1\,^\circ\text{/s}, \theta_\text{el}=60^\circ\) and \(D=1\,\mathrm{m}\).

\subsection*{Gyroscopic force} 
Here we discuss the gyroscopic effect that exists when both the rotation of the rotor and the telescope scan are present. Since the centripetal force is small, we ignore it in this subsection, and the bearing force is assumed to act vertically upward, along the \(+z\) direction.  

The angular momentum of the rotor, which is denoted by \(\bm{L}\), consists of angular momentum of its own rotation and center-of-mass motion. Their amplitudes are $\bm{L}_\text{HWP} = I\omega \hat{n}_\text{HWP}$ and $\bm{L}_\text{scan} = I\Omega\hat{z}$, respectively. The $\hat{n}_\text{HWP}$ is a unit vector parallel to the rotation axis of the rotor. 
The conservation of angular momentum parallel to $\hat{n}_\text{HWP}$ results in the scan-direction-dependent rotation speed modulation of $\omega = \overline{\omega} + \Omega \sin\theta_\text{el}$ \citep{SO_CHWP}. $\bm{L}$ is dominated by $\bm{L}_\text{HWP}$ rather than $\bm{L}_\text{scan}$.
When the telescope is stationary, \(\bm{L}\) remains constant and the net torque on the rotor is zero. In contrast, when the telescope is scanning, the direction of  \(\bm{L}\) changes and a torque of 
\begin{align}
    \frac{d\bm{L}}{dt} = I\omega\Omega\cos\theta_\text{el}\hat{y}
\end{align}
is exerted on the rotor in the tangential direction of the scan.

\begin{figure}
    \centering
    \includegraphics[width = 0.48\textwidth]{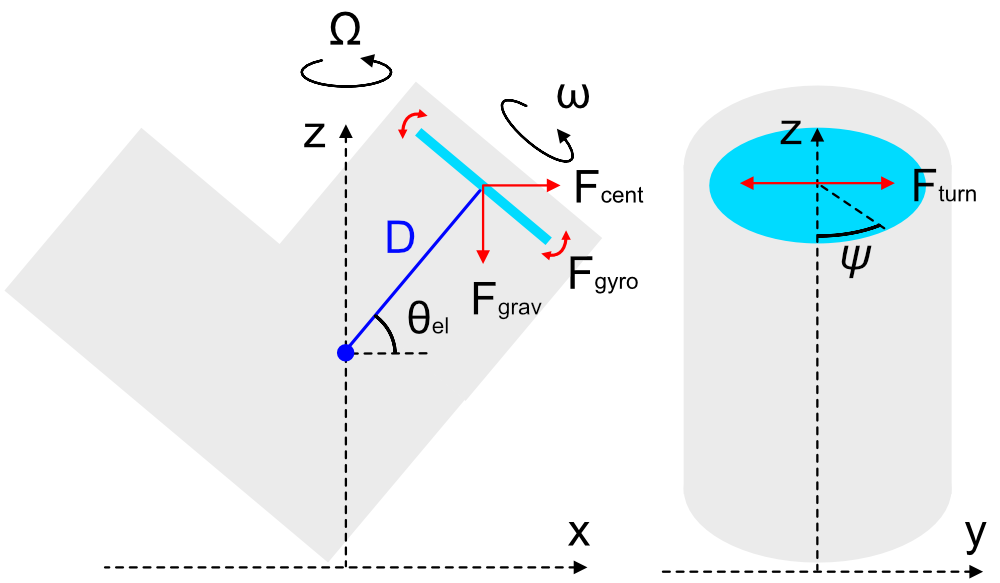}
    \caption{Schematic figure of the various forces acting on the rotor. The gray area represents the telescope and the blue disk represents the rotor.}
    \label{fig:forces}
\end{figure}

This torque arises from the non-uniform distribution of the bearing force. To discuss this, we parameterize the angular coordinate of the ring-shaped bearing by \(\psi\). The origin of \(\psi\) is defined at the lowest point of the ring, as shown in Fig.~\ref{fig:forces}. We denote the force exerted by an infinitesimal segment \(d\psi\) of the bearing as \(\bm{f}(\psi)\,r_\text{SMB}d\psi\), where \(r_\text{SMB}\) denotes the bearing radius (Fig.~\ref{fig:method} (b)). When \(\bm{f}(\psi)\) is constant, the net torque is zero, and considering the balance with gravity, we have \(\bm{f}(\psi)=\bm{F}_\text{grav}/2\pi r_\text{SMB}\). 
 Assuming \(\bm{f}_\text{gyro}(\psi)\propto\cos\psi\,\hat{n}_\text{HWP}\), we write \(\bm{f}(\psi)=\bm{F}_\text{grav}/2\pi r_\text{SMB}+\bm{f}_\text{gyro}(\psi)\). Adjusting the coefficient to reproduce the torque derived above, we obtain
\begin{align}
    \bm{f}_\text{gyro}(\psi) = -\frac{I\omega\Omega\cos\theta_\text{el}}{\pi r_\text{SMB}}\cos\psi\,\hat{n}_\text{HWP}.
\end{align}

Physically, this force distribution is realized through a slight tilt of the rotor against the stator. Given that the spring constant per unit length of the bearing for displacements parallel to the telescope axis is on the order of \(10^{4}\,\mathrm{N/m}\) \citep{SO_CHWP, Sasaki_2025}, and substituting representative values for the other parameters, the tilt is estimated to be on the order of \(0.01^\circ\).

\subsection*{Turnaround acceleration}
During telescope turnarounds, an angular acceleration \(\alpha_\text{az}\) of the telescope’s azimuth angle is applied. Approximating the center-of-mass motion of the rotor as a uniformly accelerated linear motion, a horizontal force of magnitude
\begin{align}
    \bm{F}_\text{turn} = m D\alpha_\text{az}\cos\theta_\text{el}\,\hat{y}
\end{align}
acts on the rotor. At the exact moment of turnaround, both the centrifugal force and the gyroscopic force vanish. After the turnaround, the centrifugal force returns to its original value, while the direction of the gyroscopic force reverses. The magnitude of \(\bm{F}_\text{turn}\) is approximately 0.1\% of \(\bm{F}_\text{grav}\), with representative parameters of \(\alpha=1\,^\circ \text{/s}^2, \theta_\text{el}=60^\circ\) and \(D=1\,\mathrm{m}\).

\bibliography{reference}{}
\bibliographystyle{aasjournalv7}

\end{document}